\documentclass[twocolumn, iop]{openjournal}

\usepackage{xcolor}
\usepackage{textgreek}
\usepackage[utf8]{inputenc}
\usepackage[english]{babel}

\usepackage{hyperref}
\hypersetup{
    unicode, 
    colorlinks=true,
    linkcolor=linkcolor,
    citecolor=linkcolor,
    filecolor=linkcolor,
    urlcolor=linkcolor,
}
\usepackage{color,colortbl}
\definecolor{linkcolor}{rgb}{0.0,0.3,0.5}
\usepackage{tensind}
\tensordelimiter{?}
\DeclareGraphicsExtensions{.bmp,.png,.jpg,.pdf}
\usepackage{verbatim}
\usepackage[normalem]{ulem}
\usepackage{orcidlink}
\usepackage{soul}

\usepackage{tabularx}
\usepackage{array}
\usepackage{float}

\graphicspath{ {./figs/} }

\newcommand{\lya}{\ifmmode\mathrm{Ly}\alpha\else{}Ly$\alpha$\fi\:}
\newcommand{\lyb}{\ifmmode\mathrm{Ly}\beta\else{}Ly$\beta$\fi}
\newcommand{\igm}{\ifmmode\mathrm{IGM}\else{}IGM\fi}
\newcommand{\lae}{\ifmmode\mathrm{LAE}\else{}LAE\fi}
\newcommand{\h}{\ifmmode\mathrm{H}\else{}H\fi}
\newcommand{\hi}{\ifmmode\mathrm{H\,{\scriptscriptstyle I}}\else{}H\,{\scriptsize I}\:\fi}
\newcommand{\heii}{\ifmmode\mathrm{He\,{\scriptscriptstyle II}}\else{}He\,{\scriptsize II}\fi}
\newcommand{\heiii}{\ifmmode\mathrm{He\,{\scriptscriptstyle III}}\else{}He\,{\scriptsize III}\fi}
\newcommand{\ciii}{\ifmmode\mathrm{C\,{\scriptscriptstyle III]}}\else{}C\,{\scriptsize III]}\fi}
\newcommand{\oiii}{\ifmmode\mathrm{O\,{\scriptscriptstyle III}}\else{}O\,{\scriptsize III}\fi}
\newcommand{\aliii}{\ifmmode\mathrm{Al\,{\scriptscriptstyle III}}\else{}Al\,{\scriptsize III}\fi}
\newcommand{\mgii}{\ifmmode\mathrm{Mg\,{\scriptscriptstyle II}}\else{}Mg\,{\scriptsize II}\fi}
\newcommand{\fe}{\ifmmode\mathrm{Fe}\else{}Fe\fi}
\newcommand{\nv}{\ifmmode\mathrm{N\,{\scriptscriptstyle V}}\else{}N\,{\scriptsize V}\fi}
\newcommand{\niv}{\ifmmode\mathrm{N\,{\scriptscriptstyle IV]}}\else{}N\,{\scriptsize IV]}\fi}
\newcommand{\cii}{\ifmmode\mathrm{C\,{\scriptscriptstyle II}}\else{}C\,{\scriptsize II}\fi}
\newcommand{\civ}{\ifmmode\mathrm{C\,{\scriptscriptstyle IV}}\else{}C\,{\scriptsize IV}\fi}
\newcommand{\siv}{\ifmmode\mathrm{Si\,{\scriptscriptstyle IV}}\else{}Si\,{\scriptsize IV}\fi}
\newcommand{\siii}{\ifmmode\mathrm{Si\,{\scriptscriptstyle II}}\else{}Si\,{\scriptsize II}\fi}
\newcommand{\siiii}{\ifmmode\mathrm{Si\,{\scriptscriptstyle III]}}\else{}Si\,{\scriptsize III]}\fi}
\newcommand{\ovi}{\ifmmode\mathrm{O\,{\scriptscriptstyle VI}}\else{}O\,{\scriptsize VI}\fi}
\newcommand{\sioiv}{\ifmmode\mathrm{Si\,{\scriptscriptstyle IV}\,\plus O\,{\scriptscriptstyle IV]}}\else{}Si\,{\scriptsize IV}\,+O\,{\scriptsize IV]}\fi}
\newcommand{\Msun}{\rm M_\odot}

\newcommand{\appropto}{\mathrel{\vcentre{
			\offinterlineskip\halign{\hfil$##$\cr
				\propto\cr\noalign{\kern2pt}\sim\cr\noalign{\kern-2pt}}}}}

\newcommand{\lsim}{\mathrel{\rlap{\lower4pt\hbox{\hskip1pt$\sim$}}
        \raise1pt\hbox{$<$}}}
\newcommand{\gsim}{\mathrel{\rlap{\lower4pt\hbox{\hskip1pt$\sim$}}
        \raise1pt\hbox{$>$}}}
\newcommand{\Rom}[1]{\uppercase\expandafter{\romannumeral #1}}
\newcommand{\rom}[1]{\lowercase\expandafter{\romannumeral #1}}

\newcommand{\oneh}{h^{-1}}

\newcommand{\meraxes}{\textsc{Meraxes}\:}

\begin{document}

\title{Super-Resolved Merger Trees for Cosmic Dawn: Recovering Halo Assembly and Early Galaxy Populations with Machine Learning \vspace{-1cm}}
\author{John Brennan$^{1,2,*}$\orcidlink{0000-0002-4428-6798}}
\author{Sreedhar Balu$^{3}$\orcidlink{0000-0002-5281-5151}}
\author{Yuxiang Qin$^{4}$\orcidlink{0000-0002-4314-1810}}
\author{John Regan$^2$\orcidlink{0000-0001-9072-6427}}
\author{Chris Power$^{5}$\orcidlink{0000-0002-4003-0904}}
\thanks{$^*$E-mail:john.brennan@mu.ie}
\affiliation{
$^1$NSF-Simons AI Institute for Cosmic Origins, University of Texas at Austin, Austin, TX, USA\\
$^2$Centre for Astrophysics and Space Science Maynooth, Department of Physics, Maynooth University, Maynooth, Ireland \\
$^{3}$Facultad de Físicas, Multidisciplinary Unit for Energy Science, Universidad de Sevilla, 41012, Seville, Spain\\
$^4$Research School of Astronomy and Astrophysics, Australian National University, Canberra, ACT 2611, Australia\\
$^5$International Centre for Radio Astronomy Research (ICRAR), M468, University of Western Australia, 35 Stirling Hwy, Crawley, WA 6009, Australia
}

\begin{abstract}
\noindent Machine-learning super-resolution techniques offer a route to generating high-resolution cosmological structure without the full cost of a direct high-resolution simulation. Its usefulness for (semi-analytical) galaxy-formation modelling, however, depends on recovering halo merger histories and internal properties in addition to standard dark-matter summary statistics. We test this using low-resolution, high-resolution, and super-resolution realisations of the same cosmological volume. We identify haloes, construct merger trees, and apply the \meraxes semi-analytic model which models both Pop.~III and metal-enriched Pop.~II star formation. The super-resolved simulation closely reproduce the high-resolution halo abundance and yields broadly consistent stellar mass functions for both populations across cosmic dawn. Its merger trees recover the main progenitors and massive mergers, with the remaining deficit concentrated among very low-mass progenitor branches. The principal residual in the galaxy population is a modest deficit at later times, associated with systematically larger halo spins in the super-resolution data. In \meraxes this produces more extended gas discs, lowers their surface densities and star-formation efficiency, and makes the star-formation threshold more difficult to reach. Mass-binned particle profiles and projected stacks indicate that, particularly at the low-mass end, the additional angular momentum is associated with a more extended, coherently rotating halo envelope rather than faster rotation in the halo interior. Although the small-scale phase-space structure requires further improvement, super-resolution recovers the principal halo and galaxy population statistics at substantially lower computational cost. It therefore provides a promising route to high-effective-resolution merger trees for modelling early galaxy formation in volumes or ensembles that would be prohibitively expensive to evolve directly.
\end{abstract}


 \maketitle

\section{Introduction}

\noindent Understanding the formation of the first galaxies requires models that follow both the assembly of dark-matter haloes and the baryonic processes that regulate their growth. Collisionless \textit{N}-body simulations provide the halo populations, environments, and merger histories on which semi-analytic galaxy-formation models can evolve gas, stars, and feedback using physically motivated prescriptions \citep{Lacey1993,Baugh2006,somervillephysical2015}. This combination is particularly useful at cosmic dawn $(z\gtrsim10)$, where large halo samples and finely sampled merger histories are needed to model rapidly evolving galaxy populations. Resolving the low-mass haloes that host these galaxies over a representative cosmological volume, however, makes the underlying \textit{N}-body simulation computationally expensive and in some cases intractable.

Machine-learning super-resolution (SR) offers a possible way to reduce this cost. Rather than evolving every small-scale mode directly, an SR model generates a high-resolution phase-space realisation conditioned on a comparatively inexpensive low-resolution (LR) simulation. This technique has the potential to resolve regions of the parameter space that may otherwise not be practical to resolve. Recent applications have shown that such models can recover the matter power spectrum and halo mass function, as well as halo substructure and velocity statistics \citep{Li_2021,Ni_2021,Zhang_2024}. Among these, \citet{Zhang_2024} identified haloes in the generated outputs and constructed merger trees, finding that the principal assembly trends were recovered while some systematic SR--HR differences remained. In \citet{Brennanetal2025}, we similarly found systematic SR--HR differences in the small-scale distribution of particles where the shape of SR dark-matter haloes tended to be more elongated than their HR counterparts. These results motivate the next question: whether the resulting halo catalogues and merger trees are sufficiently accurate for downstream galaxy-formation models.

Agreement in dark-matter summary statistics does not by itself guarantee agreement in galaxy properties. Semi-analytic models respond to the full assembly history and internal properties of each halo, so modest systematic differences can propagate through gas accretion, cooling, star formation, and feedback. Halo angular momentum is especially relevant because it is commonly used to set the size, surface density, and dynamical time of galactic discs. A direct comparison of galaxy populations generated from SR and high-resolution (HR) merger trees is therefore required before SR simulations can be used as a substitute for direct \textit{N}-body simulations.

In this work, we perform this comparison using three realisations of the same $10$ $h^{-1}$ cMpc cosmological volume: a $512^3$-particle LR simulation, a direct $2048^3$-particle HR simulation, and an SR realisation with an effective resolution of $2048^3$ particles generated from the LR data. We identify haloes, construct merger trees, and apply the \meraxes semi-analytic model \citep{Mutch2016,Ventura2024MNRAS.529..628V} to each dataset. We first test the recovery of the halo mass function and then compare the predicted galaxy populations. The SR outputs closely reproduce the HR halo abundance and gives broadly consistent galaxy statistics across $10\leq z\leq20$. The principal residual is associated with systematically larger halo spins in SR. Because \meraxes uses halo spin to set the disc size, these higher spins produce larger, lower-surface-density gas discs in which star formation becomes less efficient.
Our particle-level analysis links the spin excess to small-scale spatial smearing in the SR reconstruction, with the low-mass SR population showing broader particle distributions and more mass and angular momentum at large halo radii.
Nonetheless, SR provides a promising and substantially cheaper route to high-resolution merger trees, while the spin distribution identifies a specific aspect of the generated phase-space structure that requires further improvement. This is likely to be overcome via more directed training. 

The remainder of this paper is structured as follows. Section~\ref{sec:method} describes the \textit{N}-body simulations, SR reconstruction, and \meraxes galaxy-formation model. Section~\ref{sec:results} compares the total and conditional progenitor halo mass functions, the Pop.~II and Pop.~III galaxy populations, and the residual difference in halo spin. Section~\ref{sec:spin_origin} examines the physical origin of the spin excess and its connection to spatial smearing in the SR reconstruction \citep{Brennanetal2025}. Section~\ref{sec:resources} compares the computational requirements, and we summarize our conclusions in Section~\ref{sec:conclusion}. Throughout this work, we adopt cosmological parameters consistent with the Planck 2015 results \citep{Planck2015Parameters}: $(\Omega_{\rm m},\Omega_{\Lambda},\Omega_{\rm b},h)=(0.3121,0.6879,0.0491,0.6751)$.

\section{Methodology}\label{sec:method}

Our analysis separates the construction of the cosmological datasets from the training and application of the SR model. We first describe the common simulation volume and the pipeline used to identify haloes and construct merger trees in LR, HR, and SR. We then describe how the particle data are represented for machine learning, the independent simulations used for training, and the WGAN used to enhance the LR test volume. Finally, we summarize the \meraxes galaxy-formation model applied to each set of merger trees. This separation ensures that the comparisons presented in Section~\ref{sec:results} evaluate the SR model on a cosmological volume that was not used during training and propagate its halo properties through a common galaxy-formation model.

\subsection{Dark-matter simulations and analysis}
\label{ssec:simulations}

We compare three dark-matter data sets that share the same periodic volume and large-scale initial conditions. This is achieved by using the same random phases for the Fourier modes when generating the initial conditions. All the simulations have a side length of $10$ $h^{-1}$ cMpc. The LR and HR reference simulations contain $512^3$ and $2048^3$ particles, respectively, while the SR data are generated by enhancing the LR particle distribution to an effective resolution of $2048^3$ particles, as described in Section~\ref{training_data}. The corresponding particle masses are $9.56\times10^5\,\Msun$ in LR and $1.49\times10^4\,\Msun$ in both HR and SR. We choose softening lengths $(\ell)$ of $4$ percent of the  mean inter-particle distance \citep{Power2003} which is $\ell\sim0.78\oneh$ ckpc for the LR and  $\sim0.19$ h$^{-1}$ ckpc for the HR simulations. Thus, SR targets the same mass resolution as HR while replacing the direct evolution of 64 times as many particles with a learned enhancement of the LR simulation. 

We generate the initial conditions using \textsc{monofonIC} \citep{MONOFONIC:3LPT}. The dark-matter-only simulations are evolved from $z=99$ to $z=10$ using \textsc{SWIFT} \citep{Schaller_2024}. We store 76 snapshots between $z=30$ and $z=10$,  spaced evenly in cosmic time.
The comparisons presented below use six representative outputs spanning $z=20$ to $z=10$. These sample the onset and subsequent build-up of the first galaxy populations without duplicating the information in closely spaced adjacent snapshots, and lie within the regime targeted by the high-redshift implementation of \meraxes.

We process LR, HR, and SR snapshots with the same halo-analysis pipeline. Haloes are identified in each snapshot with the \textsc{VELOCIraptor} \citep{Elahietal2019} structure-finder, and their descendants and progenitors are linked with \textsc{TreeFrog} \citep{Elahietal2019TreeFrog}. These catalogues and merger trees are then supplied to \meraxes to model the corresponding galaxy populations. The generated small-scale structure in SR is stochastic, and individual low-mass SR and HR haloes therefore cannot in general be matched one-to-one. Unless otherwise stated, we consequently compare ensemble statistics at fixed halo mass rather than individual objects.

\subsection{Super-Resolution}
\label{training_data}

\begin{figure*}[t]
    \centering
    \includegraphics[width=\textwidth]{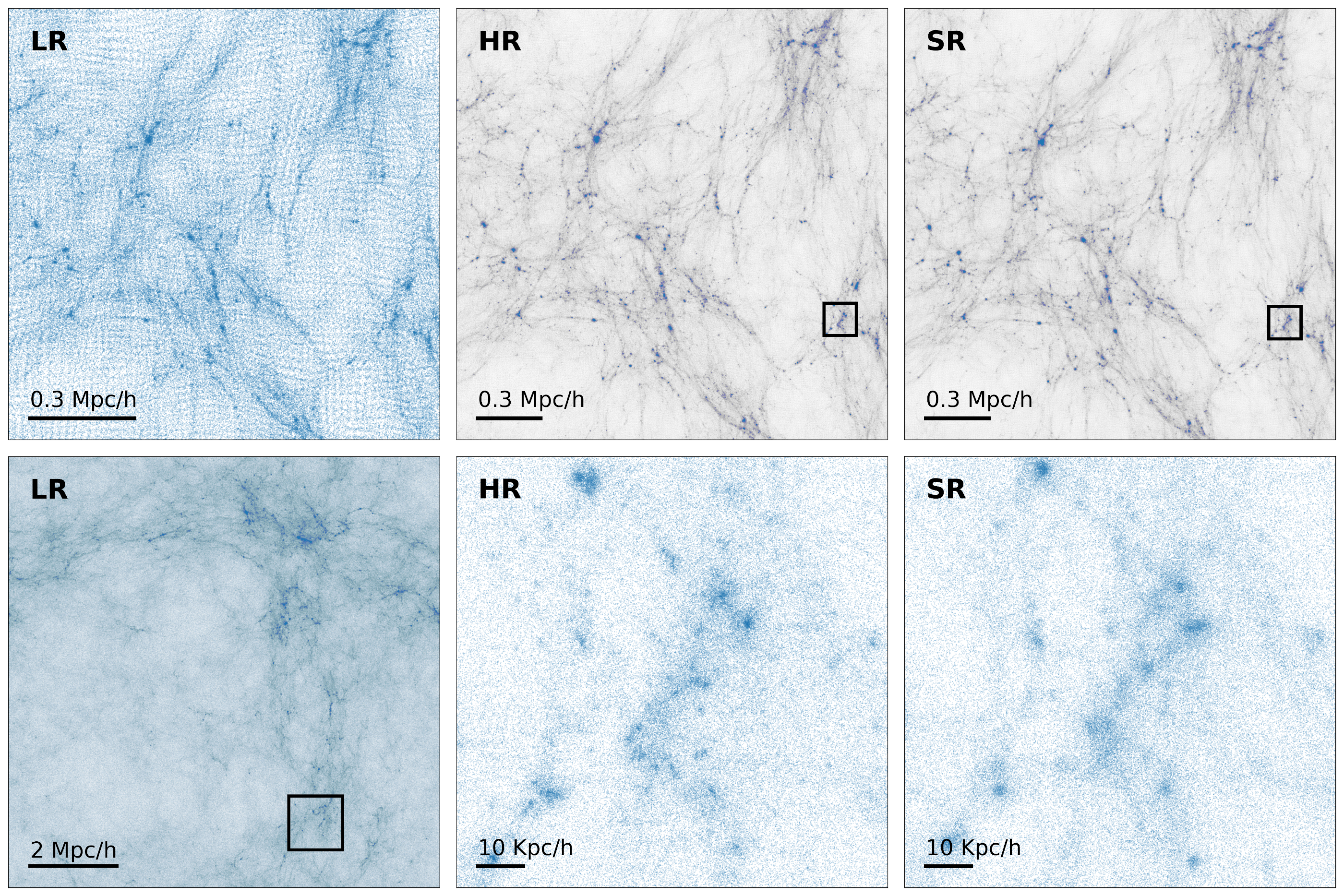}
    \caption{\textit{Illustrating halo/structure smearing on small scales within the SR data}. The bottom-left panel shows the LR particle distribution over the entire $10\,h^{-1}$ cMpc volume at redshift 10. Both HR and SR boxes have similar distributions (not shown). Within the LR volume, a $1.25\,h^{-1}$ cMpc cubic subregion (black square) is selected and highlighted in the upper row. The top row shows the particle distribution within this subregion of the LR, HR and SR snapshots. The higher  mass resolution of HR and SR w.r.t LR is apparent. We further select the same $90\,h^{-1}$ kpc subregion from HR and SR and zoom in on it in the corresponding panels in the bottom row. Even though there is broad agreement between HR and SR, the existence of smearing of structures at this spatial scale within SR is apparent.}
    \label{fig:particle_scatter}
\end{figure*}

Enhancing the particle resolution of an individual snapshot involves converting particle data to a format that is amenable to super-resolution models. To this end, we store the data of all $N^3$ particles in a snapshot in a 4D tensor with shape $6\times N\times N\times N$. We refer to this data structure as a particle tensor which is analogous to a 3D “image” where each voxel contains 6 channels: 3 components of particle displacements and 3 components of velocity. Given that the initial positions of all particles form a regular $N^3$ grid that spans the simulation box, there is a natural mapping between particles and voxels. Furthermore, since \textsc{monofonIC} assigns IDs to particles based on a ZYX-ordering (column-major) of this grid, the mapping between particle ID ($I$) and voxel indices $(i_x, i_y, i_z)$ can be made explicit as follows:
\[
    I = i_z + N i_y + N^2 i_x.
\]
To populate the particle tensor with particle data, we first invert the above equation to determine the voxel indices for each particle. Then, the displacement of each particle is computed with respect to the centre of its corresponding voxel and stored in the first three channels of the voxel. Likewise, the velocity components are stored in the final three channels. Both the LR inputs and HR targets contain all six channels, so the generator predicts particle displacements and peculiar velocities jointly; velocities are not assigned to the SR particles in a separate post-processing step. Increasing the particle resolution is then equivalent to increasing the number of voxels in the particle tensor. To be more precise, increasing the number of particles from $N^3$ to $(s\times N)^3$ requires processing the $6\times N^3$ low resolution particle tensor to generate a $6\times (s\times N)^3$ high resolution version of it. In this work, we take $s=4$, resulting in a $64$ fold increase in the number of particles. Larger augmentations are also feasible \citep[e.g.][]{Zhang_2024} but come at a higher computational cost due to increased training times. 

The super-resolution model we employ to do this is designed to operate on particle tensors of a fixed size. Namely, it takes a particle tensor of size $6\times 20^3$ as input and outputs a new particle tensor with shape $6\times 64^3$. The output tensor is taken to be a high resolution version of a central block, with shape $6\times 16^3$, within the input particle tensor. This padded configuration for the LR input is necessary as the model is based on convolutional operations that require the additional surrounding context. Hence, to enhance a full $512^3$ LR snapshot, we cut its particle tensor into $32^3$ overlapping subtensors with shape $6\times 20^3$, whose inner $16^3$ blocks cover the simulation box and do not overlap. The model then enhances each subtensor independently with the resulting output tensors being appropriately stacked together to form a final $6\times 2048^3$ particle tensor.
    
We use a Wasserstein generative adversarial neural networks model (WGAN) as our super-resolution model, whose architecture is described in detail by \citet[][see also \citealt{Brennanetal2025}]{Zhang_2024}\footnote{Our implementation of the model and the training script used to train it can be found at \url{https://github.com/brenjohn/DMSR-WGAN}}. To give a brief overview, the WGAN consists of two neural network models: A generator model and a critic model. The critic model takes HR (or SR) data as input and outputs a single numerical score for it. During training, the critic model is tuned to assign low scores to SR data and high scores to HR, thereby distinguishing between real and fake data. Meanwhile, the generator model is trained to process any LR data it is fed as input and produce fake SR data that the critic struggles to classify as fake. Once the WGAN is trained, only the generator model is used to enhance LR data as discussed above.

The generator model uses a combination of trilinear interpolation and convolutional operations to increase the spatial resolution of particle tensors. In parallel, auxiliary convolutional layers are used to transform Gaussian noise maps into small scale features before adding them to the enhanced data. The critic model, on the other hand, down-scales any HR data passed to it into a number of feature maps before computing a score for it. It does this using convolutional layers with residual connections and trilinear interpolation. To help the critic model distinguish between real and fake data, and in turn help the generator close the gap between the two, a cloud-in-cells density field is computed during training and passed to the critic alongside the HR/SR data and linearly up-scaled LR data. The cosmological scale factor of the particle data is used to modulate the weights of each convolutional layer using adaptive instance normalization \citep{Karras_2019}. This mechanism allows the model to generate enhanced data with a dependence on redshift. 
    
Training data for the WGAN model were taken from 8 pairs of LR and HR simulations of a $1.25$ $h^{-1}$ cMpc box, with $64^3$ and $256^3$ particles, respectively. A validation set was taken from a separate simulation pair with the same configuration. The test set, used for the comparative analysis, consists of a single simulation pair of a $10$ $h^{-1}$ cMpc box with $512^3$ and $2048^3$ particles in the LR and HR simulations respectively. The difference in mass resolution between LR and HR data for both the training data and the test simulations is the same: LR particles have a mass of $9.56 \times 10^{5} M_\sun$ while HR particles have a mass of $1.49 \times 10^{4} M_\sun$. From the 8 simulation pairs, each providing 100 snapshots between redshifts 35 and 5, a total of 51,200 samples of particle tensors were created for the training data. Similarly, 6,400 particle tensors were created for the validation set. Training data were augmented during training using the full symmetry group of a cube by randomly permuting and reversing the spatial dimensions of the data. Following \cite{Zhang_2024}, the model was trained using a standard WGAN loss with gradient penalty for 450 epochs using the ADAM optimizer with hyper-parameters $\beta_1 = 0$, $\beta_2=0.99$ and learning rates $l_G=1\times 10^{-5}$ and $l_C=2\times 10^{-5}$ for the generator and critic models, respectively.

Figure~\ref{fig:particle_scatter} illustrates the resulting particle distributions across the LR, HR, and SR data sets on progressively smaller spatial scales. The lower-left panel first locates a $1.25\,\oneh$ cMpc subregion within the full LR test volume, and the upper row compares this same subregion in LR, HR, and SR. Overall, SR closely reproduce the HR particle distribution, including the large-scale filaments and prominent overdensities, while recovering substantially more small-scale structure than LR. The $90\,\oneh$ kpc zoom in the lower row reveals that compact structures in SR are more diffuse than their HR counterparts, indicating spatial smearing on these scales. We examine the mass dependence of this residual and its connection to the halo-spin excess in Section~\ref{sec:spin_origin}.


\subsection{The \meraxes galaxy-formation model}
\label{ssec:meaxes}

We use the semi-analytic model \meraxes\footnote{\url{https://github.com/meraxes-devs/meraxes}} to quantify how differences between the LR, HR, and SR halo populations propagate into predictions for the first galaxies. \meraxes was developed to model galaxy formation during cosmic dawn and reionization \citep{Mutch2016}. Here, we use the extension of \citet{Ventura2024MNRAS.529..628V}, which incorporated Population~III (Pop.~III) star formation in minihaloes. We apply the same model parameters to all three sets of merger trees so that differences in the resulting galaxy populations arise solely from the underlying halo catalogues and assembly histories.

At each snapshot, \meraxes evolves the baryonic reservoirs associated with every halo. Newly accreted baryons are added to a hot-gas component, with the initial supply set by the cosmic baryon fraction. Gas then cools onto a rotationally supported cold disc through atomic cooling in sufficiently massive haloes and molecular-hydrogen cooling in minihaloes $(M_{\rm halo}\lesssim 10^7\oneh\Msun)$, with the latter also regulated by the baryon--dark-matter streaming velocity and photodissociation from a self-consistently calculated Lyman--Werner background.

Stars form from this cold-gas disc, while \meraxes follows both internal enrichment by a galaxy's own supernovae and external enrichment by expanding, supernova-driven metal bubbles. The gas metallicity determines whether newly formed stars are Pop.~III or Pop.~II, with the two populations allowed to differ in their star-formation efficiencies, surface-density thresholds, initial mass functions, and stellar yields. Supernova feedback accounts for both prompt explosions occurring within the snapshot of star formation and delayed events from longer-lived stars. We refer interested readers to \citet{Ventura2024MNRAS.529..628V} for a full description of the model and its implementation. Below, we highlight the aspects of the star-formation prescription most relevant to this work.

\begin{figure*}
    \centering
    \includegraphics[width=0.94\textwidth]{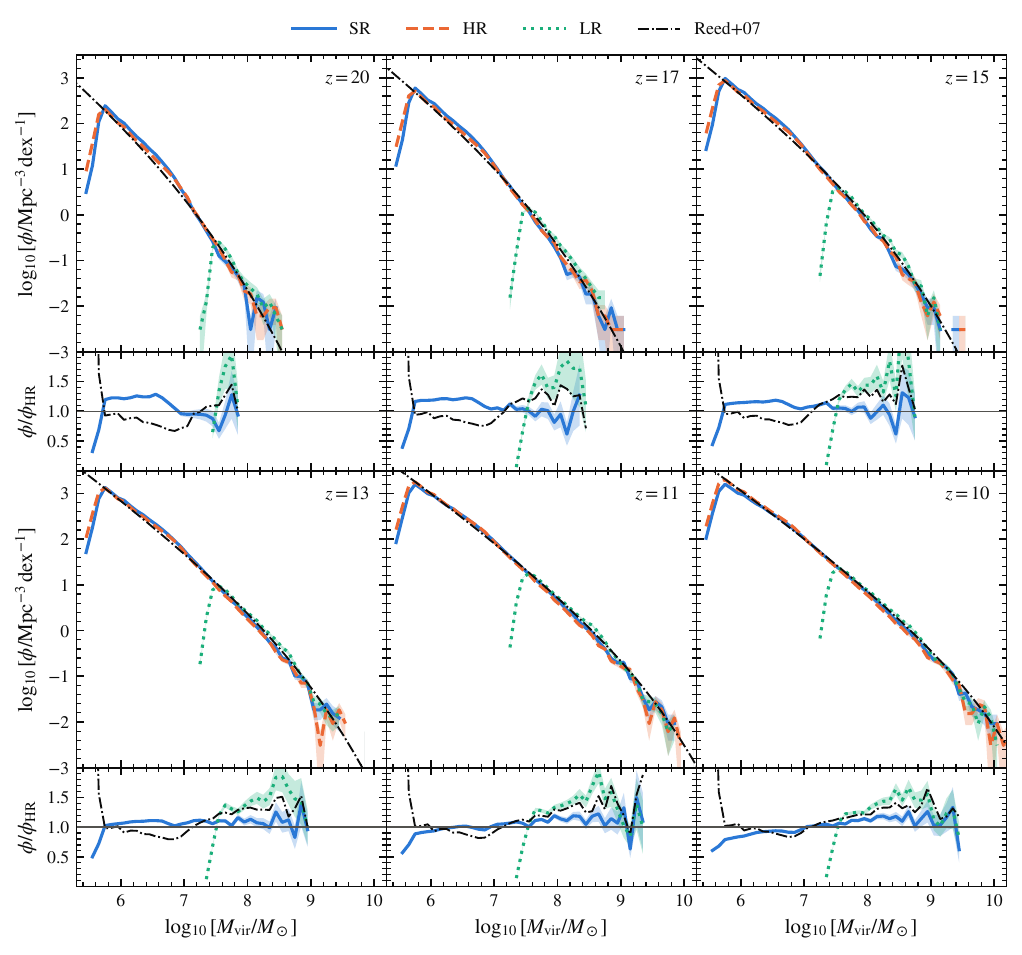}
    \caption{Halo mass functions in the SR (solid blue), HR (dashed orange), and LR (dotted green) data sets at $z=20$, 17, 15, 13, 11, and 10. The high-redshift fitting function of \citet{Reed2007} is shown for reference (black dot-dashed). For each redshift, the main panel gives the comoving halo abundance per logarithmic mass interval, while the panel beneath shows the ratio of each curve to the HR mass function. Shaded regions show $1\sigma$ Poisson uncertainties, propagated between the two simulation samples in the ratio panels. 
    Ratios are computed only for mass bins with at least 10 haloes.
    }
    \label{fig:hmfs}
\end{figure*}

The size of the cold-gas disc depends directly on the halo angular momentum. Its exponential scale radius is
\begin{equation}
  r_{\rm s} = R_{\rm vir}\frac{\lambda}{\sqrt{2}},
  \label{eq:scale_radius}
\end{equation}
where $\lambda$ is the halo spin parameter  \citep[as defined in][]{Bullock2001ApJ...555..240B}, and $R_{\rm vir}$ is the virial radius. Star formation is triggered once the cold-gas mass exceeds the critical value given by
\begin{equation}
  m_{\rm crit} = 2\pi\Sigma_{\rm norm}
  \left(\frac{V_{\rm vir}}{{\rm km\,s^{-1}}}\right)
  \left(\frac{r_{\rm disc}}{{\rm kpc}}\right)10^6\,\Msun,
  \label{eq:critical_mass}
\end{equation}
where $r_{\rm disc}=3r_{\rm s}$ and $\Sigma_{\rm norm}$ is a model parameter that sets the critical gas surface density. For $m_{\rm cold}>m_{\rm crit}$, the star-formation rate is
\begin{equation}
  \dot{m}_* = \alpha_{\rm SF}
  \frac{m_{\rm cold}-m_{\rm crit}}{t_{\rm dyn}^{\rm disc}},
  \label{eq:star_formation_rate}
\end{equation}
where $\alpha_{\rm SF}$ is the star-formation efficiency and $t_{\rm dyn}^{\rm disc}=r_{\rm disc}/V_{\rm vir}$ is the disc dynamical time. Equations~\ref{eq:scale_radius}--\ref{eq:star_formation_rate} make the galaxy properties sensitive to systematic differences in halo spin: at fixed cold-gas mass and virial properties, a larger $\lambda$ produces a more extended disc, hence a larger threshold mass for star formation, and a lower star-formation rate once the threshold is exceeded.
Relating disc size to halo angular momentum is a standard assumption in many semi-analytic models \citep{somervillephysical2015}. Here we retain the prescription adopted by 
\cite{Ventura2024MNRAS.529..628V} and apply it identically to all three merger-tree sets; the comparison therefore isolates how differences in the input halo spins propagate through a common modelling framework.


\section{Results}\label{sec:results}

We first test whether SR reproduces the abundance of dark-matter haloes before introducing any galaxy-formation physics. We then use conditional progenitor halo mass functions to test whether the SR merger trees recover the assembly histories relevant to early star formation, before comparing the resulting galaxy stellar mass functions. This progression separates differences in the underlying halo abundance from those introduced by halo assembly histories or internal halo properties.

\subsection{Halo mass functions}

\begin{figure*}[ht]
    \centering
    \includegraphics[width=\textwidth]{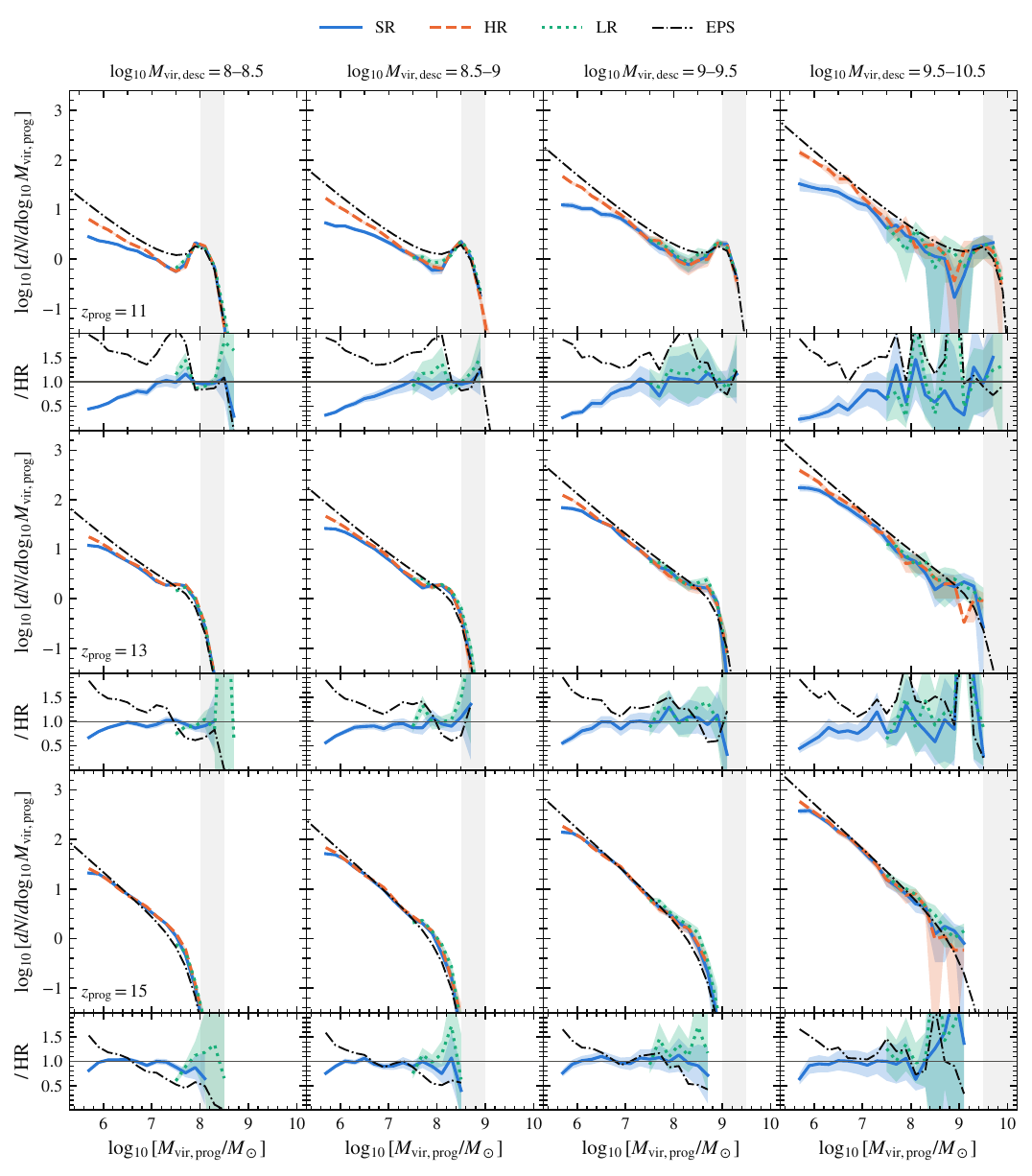}
    \caption{Conditional (progenitor) halo mass functions for descendants at $z=10$. Columns select host haloes in the descendant total-mass intervals $\log_{10}(M_{\rm vir,desc}/\Msun)=8$--$8.5$, $8.5$--$9$, $9$--$9.5$, and $9.5$--$10.5$ (grey bands), while rows show their progenitors at $z_{\rm prog}=11$, 13, and 15. The main panels give the mean number of distinct progenitors per descendant and per logarithmic progenitor-mass interval for SR (solid blue), HR (dashed orange), and LR (dotted green). The extended Press--Schechter prediction is shown for reference (EPS; black dot-dashed; \citealt{Bond1991,Lacey1993}). Within each column, the three simulation samples are re-weighted to a common descendant total-mass distribution. Shaded regions in the main panels show 95 per cent bootstrap intervals obtained by resampling descendants. The smaller panels show each curve relative to HR, with approximate 95 per cent intervals propagated from the independent bootstrap uncertainties of the simulation samples.}
    \label{fig:conditional_hmfs}
\end{figure*}

Figure~\ref{fig:hmfs} shows the recovery of the total halo population independently of the baryonic prescriptions in \meraxes. All bands show $1\sigma$ Poisson counting uncertainties. The lowest plotted halo masses correspond approximately to the 20-particle scale, $3.0\times10^5\,\Msun$ in SR and HR and $1.9\times10^7\,\Msun$ in LR. We conservatively restrict the quantitative SR-HR comparison to $M\geq10^6\,\Msun$. Across $10^6\lesssim M/\Msun\lesssim10^8$, the SR mass function remains within approximately 25 per cent of HR at all six redshifts and closely reproduces its shape and redshift evolution. Larger excursions are confined to the resolution edge and the sparsely populated high-mass tail, where the statistical uncertainties are correspondingly larger. For reference, the fitting function of \citet{Reed2007} is broadly consistent with the simulated mass functions across their well-resolved mass range.

The LR mass function is consistent with HR only above its substantially larger resolution scale. Its abundance turns over near a few $10^7\,\Msun$ and then falls rapidly towards lower masses, whereas SR continues to trace HR by more than an order of magnitude further down the halo mass function. The agreement between SR and HR therefore demonstrates that the learned enhancement recovers not only the additional low-mass haloes absent from LR, but also their evolution across $10\leq z\leq20$. This establishes the halo population required for the subsequent merger-history and galaxy-formation tests.

\begin{figure*}
    \centering
    \includegraphics[width=0.95\textwidth]{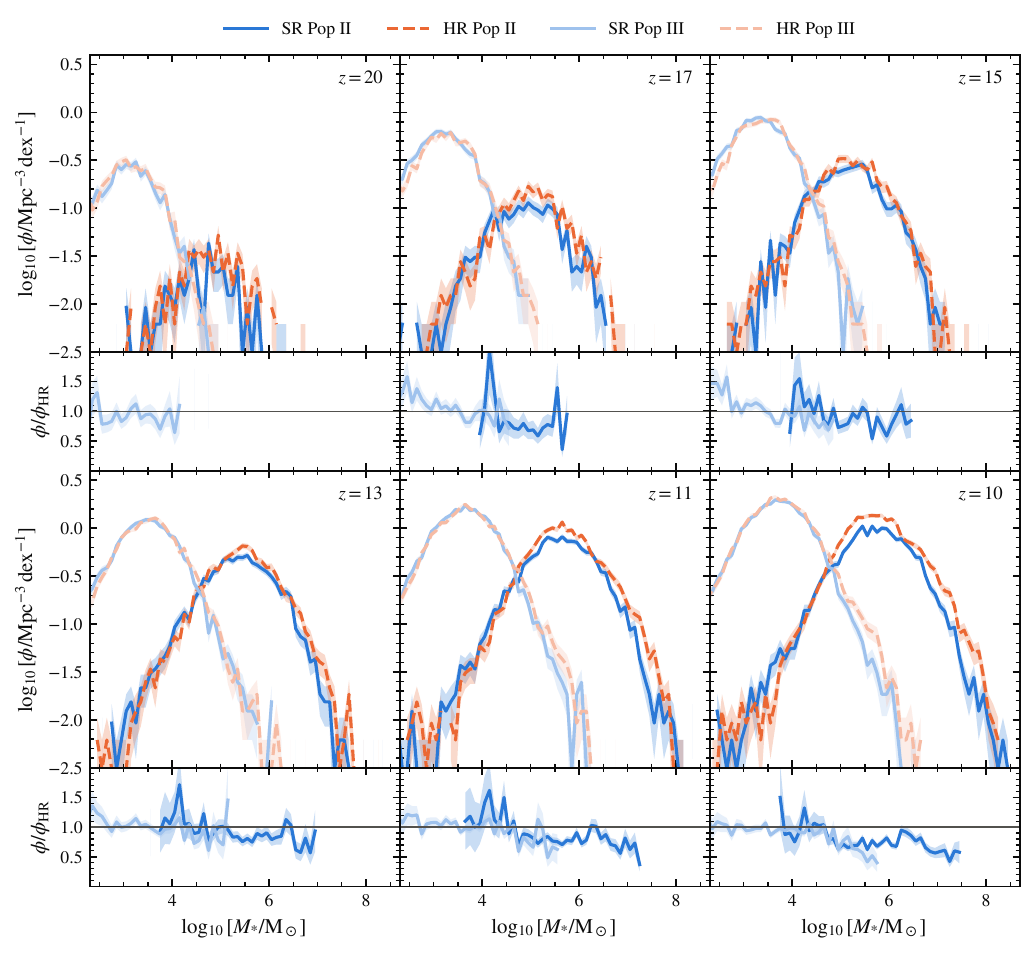}
    \caption{Galaxy stellar mass functions predicted by \meraxes from the SR (blue solid) and HR (orange dashed) merger trees at $z=20$, 17, 15, 13, 11, and 10. Dark and light curves show the Pop.~II and Pop.~III stellar components, respectively. The main panels give the comoving abundance per logarithmic stellar-mass interval, while the smaller panels show the SR-to-HR ratios. Shaded regions show $1\sigma$ Poisson uncertainties, propagated between the two samples in the ratio panels. }
    \label{fig:gsmfs}
\end{figure*}

\subsection{Conditional progenitor halo mass functions}

Agreement in the total HMFs establishes that SR recovers the number of haloes at each snapshot, but does not test whether those haloes are connected through time with the correct growth histories. Figure~\ref{fig:conditional_hmfs} therefore shows the conditional, or progenitor, halo mass functions of descendants at $z=10$, using $M_{\rm vir}$ for both descendants and progenitors. We select host haloes in four descendant total-mass intervals and trace all host-halo progenitors associated with their final top-level FOF groups at $z_{\rm prog}=11$, 13, and 15. For each sample, we calculate the mean number of progenitors per descendant and per dex in progenitor mass. To ensure a like-for-like comparison, we divide each descendant-mass interval into $0.05$-dex sub-bins and weight the three simulations to contain the same effective number of descendants in each sub-bin. Uncertainties are obtained by bootstrap-resampling descendants, thereby preserving correlations among progenitors of the same halo. The highest-mass interval is sample-limited and has correspondingly broad uncertainties. The EPS prediction is also included as a theoretical reference \citep{Bond1991,Lacey1993}.

At $z_{\rm prog}=11$, the conditional mass functions contain a narrow main-progenitor peak immediately below the descendant-mass interval, corresponding to approximately one main progenitor per descendant, together with a tail of lower-mass progenitors. Towards higher redshift, the typical main-progenitor mass decreases and the peak becomes progressively less distinct from the low-mass tail. Within approximately one decade of the descendant mass, SR closely follows HR within the uncertainties across all three progenitor redshifts. The main progenitors and most massive mergers that dominate halo growth are therefore well recovered. The EPS curves are also broadly consistent with the simulated conditional mass functions and reproduce their principal mass and redshift trends.

The principal SR-HR discrepancy occurs among the least massive progenitors. At $M_{\rm vir,prog}\lesssim10^7\,\Msun$, SR falls below HR most clearly at $z_{\rm prog}=11$, with the largest separation for the more massive descendants. The offset becomes progressively smaller towards $z_{\rm prog}=13$ and 15, where the SR and HR curves agree over most of the resolved mass range. Because the difference is concentrated in the low-mass tail while the main progenitors agree, it affects the number of recovered branches more strongly than the dominant mass assembly.

\begin{figure*}
    \centering
    \includegraphics[width=0.92\textwidth]{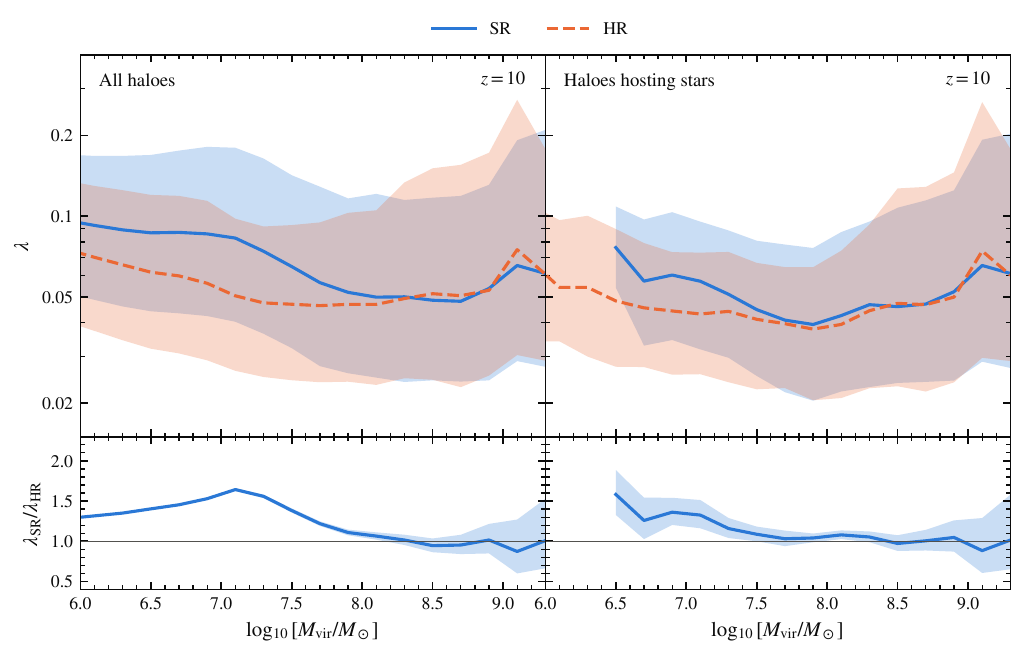}
    \caption{Halo spin parameter as a function of virial mass at $z=10$ for all haloes (left) and for haloes hosting stars in \meraxes (right). The upper panels compare the binned spin distributions in SR (solid blue) and HR (dashed orange), with shaded regions enclosing the central 95 per cent range within each mass bin. The lower panels show the ratio of the SR and HR median spin parameters. In both samples, SR has systematically higher median spins at the low-mass end, with the offset decreasing towards higher halo masses.}
    \label{fig:spin_mass}
\end{figure*}

The deficit in low-mass progenitors is more pronounced than the corresponding difference in the global halo mass function over the same mass range, suggesting that it is enhanced in the overdense regions that assemble into the selected $z=10$ descendants. Some of this enhancement may, however, reflect incomplete branch tracking rather than a difference in halo abundance. The tree statistic cannot distinguish haloes that are absent from the SR catalogues from those that are identified at high redshift but whose branches terminate before $z=10$. A direct completeness check of all \textsc{TreeFrog} branches at the three progenitor snapshots shows that approximately 88--90 per cent of SR branches reach $z=10$, compared with 94--95 per cent in HR. The additional branch attrition in SR therefore likely contributes to the conditional deficit. We interpret Figure~\ref{fig:conditional_hmfs} as demonstrating a deficit of recovered low-mass progenitor branches, rather than as establishing that the corresponding haloes are physically absent. Above its 20-particle limit, LR is consistent with HR within the statistical uncertainties, but resolves far fewer progenitors towards high redshift.

\subsection{Pop.~II and Pop.~III stellar mass functions}
\label{ssec:galaxy_results}

Figure~\ref{fig:gsmfs} compares the Pop.~II and Pop.~III stellar mass functions predicted from applying \meraxes on the SR and HR merger trees. The two components occupy distinct stellar-mass ranges: Pop.~III dominates the lower-mass peak, while Pop.~II extends to progressively larger stellar masses as galaxies grow and enrich. SR reproduces this separation, together with the shape and redshift evolution of both populations.

At $z\gtrsim15$, the SR and HR predictions generally overlap within the uncertainties across the well-sampled mass range. A small systematic difference develops towards lower redshift. By $z=13$--10, SR tends to predict a lower abundance on the high-mass side of both the Pop.~II and Pop.~III distributions, while agreement remains closer around their peaks and at lower stellar masses. The difference is largest in the rapidly declining tails, where the statistical uncertainties are also broader, but the overall population statistics remain well reproduced.

The common direction of the offset in both stellar populations suggests a difference in the efficiency with which galaxies build stellar mass, rather than a failure to recover either the Pop.~III or Pop.~II population. Because the total halo mass functions and the massive progenitor branches remain consistent between SR and HR, this residual points to more subtle differences in halo properties. We next examine halo spin, which directly regulates disc size and the star-formation threshold in \meraxes.

\subsection{Residual difference in halo spin}

Figure~\ref{fig:spin_mass} isolates one internal halo property that directly affects the \meraxes star-formation prescription: the spin. Among all haloes at $z=10$, the median spin in SR is approximately 30--60 per cent higher than in HR across $10^6\lesssim M_{\rm vir}/\Msun\lesssim10^{7.5}$, with the largest difference near $M_{\rm vir}\simeq10^7\,\Msun$. The offset then decreases with increasing mass: the median spins differ by only about 10 per cent at $M_{\rm vir}\simeq10^8\,\Msun$ and are consistent at higher masses, where the halo-to-halo scatter and sampling uncertainties also increase. The spin excess is therefore concentrated in the low-mass population rather than representing a uniform shift across all SR haloes.

The same mass dependence is present when the comparison is restricted to haloes that host stars (second column of Figure \ref{fig:spin_mass}). Their median SR-to-HR spin ratio is approximately $1.3$--$1.6$ below $M_{\rm vir}\simeq10^7\,\Msun$, but approaches unity by $M_{\rm vir}\simeq10^{7.5}\,\Msun$. A halo is included in this panel only after its cold-gas mass has crossed the star-formation threshold in Equation~\ref{eq:critical_mass}; even a relatively massive halo can therefore remain absent from the sample if it has not yet satisfied this condition. The systematic spin offset persists in the objects that contribute directly to the stellar mass functions, although the broad and overlapping distributions show that it is a population-level effect.

Through Equations~\ref{eq:scale_radius}--\ref{eq:star_formation_rate}, the larger spins produce more extended discs, lower their cold-gas surface densities, and raise the cold-gas mass required for star formation.
The resulting suppression of stellar-mass growth can accumulate over successive stages of galaxy assembly, providing a natural explanation for the lower SR abundances on the high-stellar-mass sides of the Pop.~II and Pop.~III distributions in Figure~\ref{fig:gsmfs}. 
We next examine the particle distributions of these haloes to identify the origin of this additional angular momentum.

\section{The origin of the halo-spin excess: spatial smearing}\label{sec:spin_origin}

\begin{figure*}[t]
    \centering
    \includegraphics[width=\textwidth]{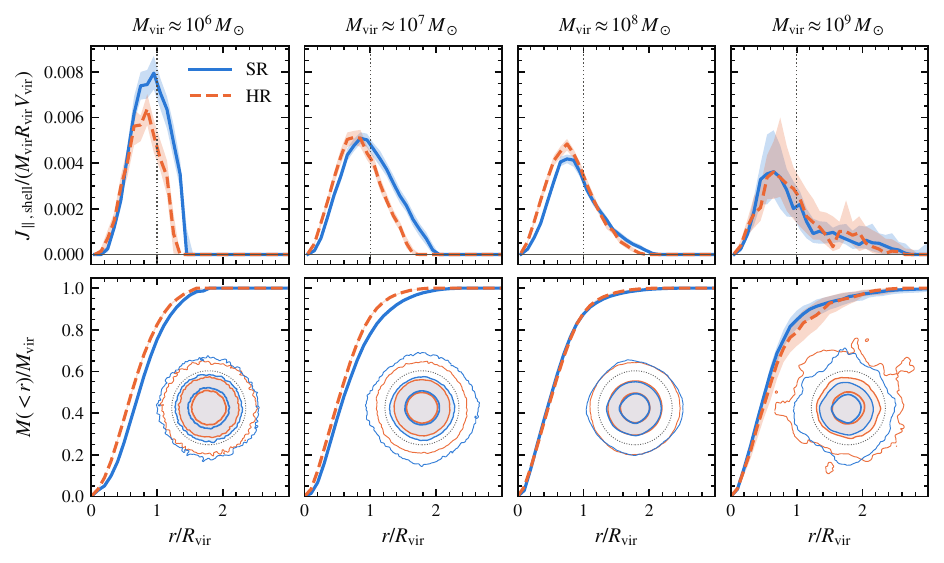}
    \caption{Radial structure of mass-matched SR (blue solid) and HR (orange dashed) haloes at $z=10$ in four total-mass bins centred on $10^6$, $10^7$, $10^8$, and $10^9\,\Msun$ (left to right). The top row shows the angular momentum in spherical shells projected along each halo's total angular-momentum direction, normalized by $M_{\rm vir}R_{\rm vir}V_{\rm vir}$. The bottom row shows the cumulative enclosed mass normalized by $M_{\rm vir}$. Shaded regions denote 95 per cent halo-bootstrap intervals on the population median. The embedded contours in the lower panels show the projected particle distributions of the same haloes after stacking. While solid contours enclose 50, 80, and 95 per cent of the projected probability, the dotted circle marks $R_{\rm vir}$.}
    \label{fig:radial_profiles}
\end{figure*}

To investigate the origin of the additional angular momentum in SR haloes, we compare the SR and HR halo populations at $z=10$ in four mass bins centred on $10^6$, $10^7$, $10^8$, and $10^9\,\Msun$. Because SR is generated stochastically, its low-mass haloes do not necessarily correspond one-to-one with those in HR. We therefore compare the two populations at fixed total mass\footnote{We randomly draw at most 2,000 objects per simulation within $0.05$ dex of each bin centre, retain those with complete radial profiles and at least 50 member particles, and use the same haloes for the profiles and stacks shown in Fig.~\ref{fig:radial_profiles}. Reweighting SR and HR in $0.01$-dex mass cells gives effective matched samples of 1855, 1944, 1840, and 67 objects per simulation from low to high mass.}.

For each selected halo, we use all member particles identified by \textsc{VELOCIraptor}, centre their positions on the potential minimum, and normalize their radii by $R_{\rm vir}$. From these particles, we calculate (i) the stacked projected particle distribution\footnote{For each halo, the particles are projected along the $xy$, $xz$, and $yz$ planes. The three maps are normalized by the halo's particle count and averaged without principal-axis alignment.}, (ii) the cumulative enclosed mass, and (iii) the angular momentum in spherical shells projected along the halo's total angular-momentum direction. Appendix~\ref{app:jshell} illustrates how this signed shell-angular-momentum statistic is constructed from individual particle contributions.

Figure~\ref{fig:radial_profiles} shows that the SR--HR difference is strongly mass dependent. At $M\simeq10^6\,\Msun$, SR haloes build up their enclosed mass more gradually than HR, while their shell-angular-momentum profile reaches a larger peak closer to $R_{\rm vir}$. A clear separation in the mass profiles remains at $M\simeq10^7\,\Msun$; the angular-momentum profiles have comparable peaks, but the SR profile extends to larger radii. The projected stacks show the same redistribution in two dimensions, with broader outer SR contours at $10^6$--$10^7\,\Msun$. By $M\simeq10^8\,\Msun$, the profiles and stacks are substantially more consistent. SR and HR also remain consistent in the $10^9\,\Msun$ bin, although this comparison is more strongly affected by sample variance.

This population-level comparison strengthens the visual evidence introduced in Figure~\ref{fig:particle_scatter}, where the 90 h$^{-1}$ kpc zoom shows compact HR structures appearing more diffuse in SR. Figure~\ref{fig:radial_profiles} demonstrates that the same behaviour persists in controlled, mass-matched ensembles: the low-mass SR population has a broader enclosed-mass profile and projected particle distribution. We describe this redistribution as \emph{spatial smearing}, consistent with the small-scale behaviour reported in our previous work \citep{Brennanetal2025}. The associated redistribution of angular momentum provides a direct structural explanation for the larger spins of low-mass SR haloes. Rather than being driven primarily by faster rotation at small radii, the excess is associated with a more extended, coherently rotating outer envelope\footnote{We also examined the azimuthal velocities, velocity-dispersion tensor, retrograde mass fraction, angular-momentum cancellation factors, and a velocity-direction scrambling test. Although SR haloes are dynamically colder, these complementary diagnostics show that neither faster rotation nor velocity dispersion alone explains the spin offset: SR haloes contain fewer counter-rotating particles and exhibit less angular-momentum cancellation, indicating more coherent rotation.}. Future work will use this information to improve the SR prediction of halo structure and phase-space correlations on these scales, while preserving its successful recovery of the halo and galaxy populations.

\section{Computational Resources}\label{sec:resources}

The current SR approach provides a compelling computational trade-off. It recovers the principal population-level statistics of the HR simulation at a significantly lower computational cost. The small-scale discrepancies identified above provide clear targets for improvement, but do not diminish the central practical advantage of replacing a prohibitively expensive direct HR \textit{N}-body simulation with a much cheaper SR workflow. We therefore compare the resources required by the production workflows used in this work.

The LR simulation required approximately $3{,}500$ CPU hours. Enhancing one LR snapshot required approximately $930$ CPU hours on a single 128-CPU node; enhancing all 76 snapshots therefore used just over $70{,}000$ CPU hours. Post-processing the resulting SR snapshots with \textsc{VELOCIraptor} required a further $23{,}000$ CPU hours, while the costs of \textsc{TreeFrog} and \meraxes were negligible in comparison. The per-volume SR production\footnote{Training the SR model required 256 GB of memory, split across 4 NVIDIA H100 GPUs on a single node, and took 192 node hours to complete. However, this cost is incurred once and can be amortised over subsequent volumes processed with the same trained model.} workflow consequently required approximately $96{,}500$ CPU hours. By contrast, the HR simulation  required approximately $2\times10^6$ CPU hours (including halo-finding in post-processing). SR therefore produced the particle data, halo catalogues, merger trees, and galaxy predictions for approximately 5 per cent of the HR computational cost; equivalently, the direct HR workflow was about 20 times more expensive. 
This reduction changes the scale of the simulations that can be attempted. It makes high-effective-resolution snapshot sequences, and hence merger-tree and galaxy-formation analyses, feasible for volumes or ensembles for which direct HR evolution would be prohibitively expensive. 

\section{Conclusions and Outlook}\label{sec:conclusion}

We have tested whether a machine-learning SR workflow can provide merger trees suitable for semi-analytic modelling of the first galaxies. We compared LR, HR, and SR realisations of the same $10$ h$^{-1}$ cMpc volume and processed each with the same halo-finding and tree-building pipeline. We then applied the \meraxes\, galaxy-formation SAM, that tracks both Pop.~III and Pop.~II star formation, on these merger trees. This provides an end-to-end test of SR in which differences in the generated dark-matter structure are propagated into predictions for the galaxy population.

Across all redshifts, SR closely reproduces the HR halo mass function over their common resolved mass range, whereas LR turns over at a higher mass, as expected from its lower particle resolution. The conditional mass functions also agree around the main-progenitor feature and for progenitors within approximately one decade of the descendant mass. SR contains fewer low-mass progenitor branches, particularly close to $z=10$, but the agreement of the massive progenitors indicates that this difference affects branch counts more strongly than the dominant mass assembly. The Pop.~II and Pop.~III stellar mass functions generated from the SR trees reproduce the characteristic mass scales, shapes, and redshift evolution of both populations. At the later outputs, SR develops a modest deficit towards the high-stellar-mass tails of both components despite the continued agreement of the total halo populations. This shows that the dominant SR result is the successful recovery of the halo and galaxy abundances, while the remaining difference arises from more subtle aspects of halo assembly or internal structure.

We identify halo spin as one such internal property. At fixed virial mass, SR haloes have systematically larger spins, which translate in \meraxes into larger gas discs, lower surface densities, and a higher cold-gas mass threshold for star formation (see Equations \ref{eq:scale_radius}-\ref{eq:star_formation_rate}). The mass-binned particle comparison at $z=10$ shows that the structural difference is strongest near $M_{\rm vir}\simeq10^6$--$10^7\,\Msun$: these SR haloes build up their enclosed mass more gradually, have broader projected particle distributions, and carry angular momentum to larger radii than HR. The radial profiles and projected stacks become substantially more similar by $M_{\rm vir}\simeq10^8\,\Msun$, while the highest-mass comparison is limited by the small number of objects in the narrow mass bin. The spin offset is therefore mass dependent and is primarily associated with an extended, coherently rotating envelope around low-mass SR haloes rather than with faster rotation in their interiors.

We describe this extended low-mass structure as spatial smearing in the SR reconstruction, consistent with the direct particle comparison in Figure~\ref{fig:particle_scatter} and with the small-scale behaviour reported in \citet{Brennanetal2025}. The trend persists when the catalogue cap and the inclusion of subhaloes are varied, although the present diagnostics do not establish a unique cause for the redistribution. Future work will use them to refine the training procedure and improve the recovery of small-scale halo structure and phase-space correlations while preserving the successful recovery of the halo and galaxy populations.

The computational advantage is substantial. After training, the complete SR production workflow used here required 
only about 5 per cent of the HR computational cost. Taken together, our results show that SR already provides a practical and efficient route to high-resolution merger trees for early-Universe modelling. 

\section{Code and Data Access}
    \noindent All neural network models and scripts used to train and enhance snapshots for this paper were implemented in Python and PyTorch \citep{Paszke2019-hd} and are available at \url{https://github.com/brenjohn/DMSR-WGAN} on the \texttt{SR-trees} branch. Steps for reproducing the model used in this work are also available there along with our model's weights. The complete dataset used by the \meraxes runs presented here are available on Zenodo (\url{https://zenodo.org/records/23293392}; \citealt{SR_meraxes}).

\section*{Acknowledgments}
    JB and JR acknowledge support from the Royal Society and Research Ireland through the University Research Fellow programme under grant number URF$\backslash$R1$\backslash$191132.
    JR acknowledges support from the Research Ireland Laureate programme under grant number IRCLA/2022/1165. This work was supported by resources provided by the Pawsey Supercomputing
    Research Centre with funding from the Australian Government and the
    Government of Western Australia.  Parts of this research was undertaken with the assistance of resources from the National Computational Infrastructure (NCI Australia), an NCRIS enabled capability supported by the Australian Government. SB acknowledges support from grant PID2022-138855NB-C32 funded by MICIU/AEI/10.13039/501100011033 and  ERDF/EU.
    YQ is supported by the ARC Discovery Early Career Researcher Award (DECRA) through fellowship \#DE240101129.
    
    We acknowledge EuroHPC Joint Undertaking for awarding us access to MareNostrum5 hosted by BSC in Spain. The research in this paper made use of the SWIFT open-source simulation code (http://www.swiftsim.com, \citealt{Schaller_2024}) version v2025.01.
    
    During the preparation of this manuscript, the authors used an AI language assistant to help refine prose clarity and LaTeX formatting. All scientific content, analyses, and conclusions were reviewed and verified by the authors, who take full responsibility for the final text.

\appendix
\setcounter{figure}{0}
\renewcommand{\thefigure}{A\arabic{figure}}
\renewcommand{\theHfigure}{appendix.A.\arabic{figure}}
\renewcommand{\theHequation}{appendix.A.\arabic{equation}}
\section{Illustration of the shell angular momentum}
\label{app:jshell}

The top panels of Figure~\ref{fig:radial_profiles} show the signed angular momentum in spherical shells of width $0.1R_{\rm vir}$. For each halo, we take the spin axis $\hat{\mathbf{J}}$ to be the direction of the total angular momentum of its member particles and measure
\begin{equation}
  \frac{J_{\parallel,{\rm shell}}}{M_{\rm vir}R_{\rm vir}V_{\rm vir}}
  = \frac{1}{M_{\rm vir}R_{\rm vir}V_{\rm vir}}
  \sum_{i\in{\rm shell}}m_iR_{\perp,i}v_{\phi,i},
  \label{eq:jshell}
\end{equation}
where $R_{\perp,i}$ is the cylindrical distance of particle $i$ from the spin axis and $v_{\phi,i}$ is its azimuthal velocity, positive in the direction of the halo spin. Particle positions are measured relative to the minimum-potential centre, while the velocities include the Hubble flow and are measured relative to the mass-weighted mean velocity of all halo members. With the same particles and reference frame, summing Equation~\ref{eq:jshell} over all shells gives $J/(M_{\rm vir}R_{\rm vir}V_{\rm vir})=\sqrt{2}\lambda$.

\begin{figure*}
    \centering
    \includegraphics[width=\textwidth]{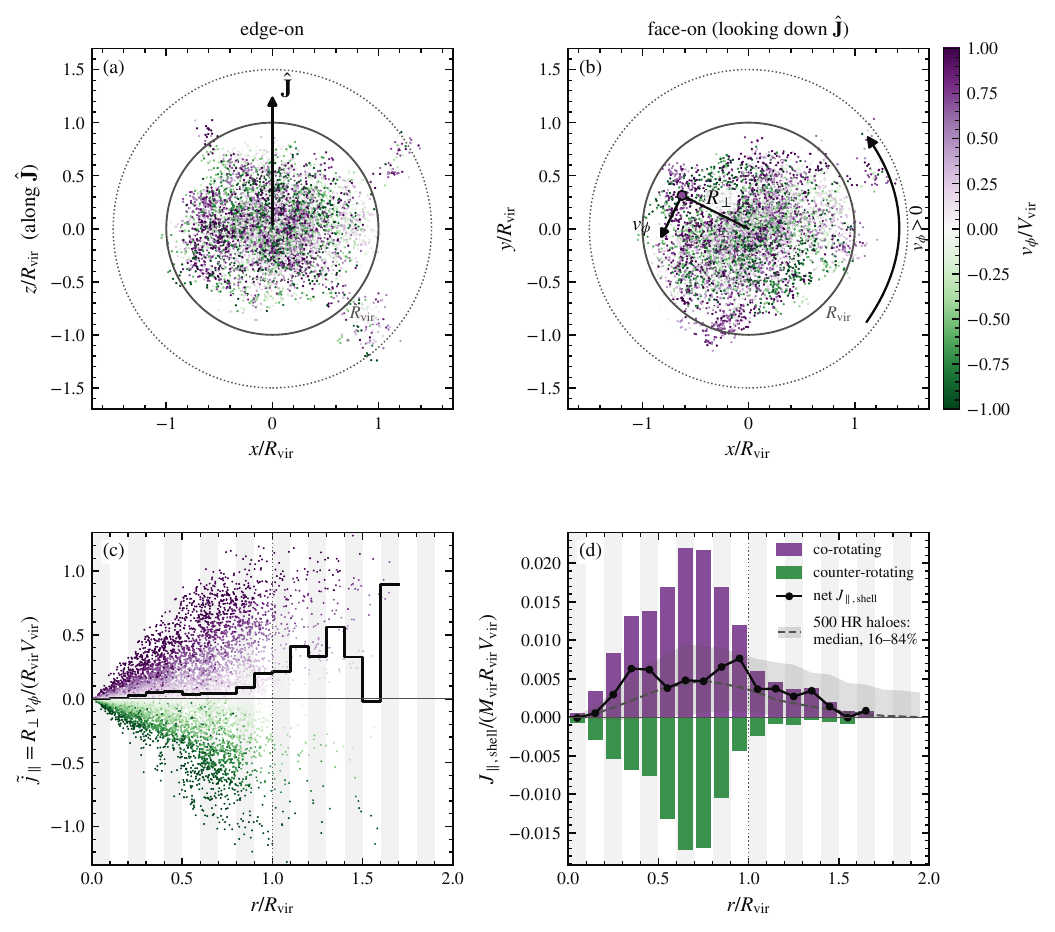}
    \caption{Construction of $J_{\parallel,{\rm shell}}$ for a representative HR halo with $M_{\rm vir}\sim10^8\,\Msun$ at $z=10$ ($7{,}335$ particles and $\lambda=0.042$). Panels (a) and (b) show edge-on and face-on projections in a frame whose $z$-axis is the spin axis $\hat{\mathbf{J}}$. Particles are coloured by $v_\phi/V_{\rm vir}$, with purple and green denoting co- and counter-rotation, respectively; circles mark $R_{\rm vir}$ and $1.5R_{\rm vir}$. The annotated particle in panel (b) contributes $mR_\perp v_\phi$ to $J_\parallel$. Panel (c) shows the dimensionless particle specific angular momentum $\widetilde{j}_\parallel=R_\perp v_\phi/(R_{\rm vir}V_{\rm vir})$; alternating shading marks the $0.1R_{\rm vir}$ shells and the black step gives the mean within each shell. Panel (d) separates the co-rotating and counter-rotating contributions to Equation~\ref{eq:jshell}, with their net shown in black. The dashed curve and grey band give the median and 16th--84th percentiles of the individual profiles of 500 HR haloes in the same mass bin. The band therefore shows halo-to-halo scatter, whereas Figure~\ref{fig:radial_profiles} shows the bootstrap uncertainty on the population median.}
    \label{fig:jshell_illustration}
\end{figure*}

We select a well-resolved halo from HR with $M_{\rm vir}\sim10^8\,\Msun$ and illustrate its particle distribution and contribution to the total angular momentum in Figure~\ref{fig:jshell_illustration}. See the caption for more details. The individual particle contributions in panel (c) are much larger than their net within a shell because co- and counter-rotating material are spatially mixed. Across the full example halo, the co-rotating and counter-rotating contributions are $+0.151$ and $-0.092$, respectively, leaving a net of $0.059=\sqrt{2}\lambda$; 44 per cent of the mass is counter-rotating. At $0.5$--$0.8R_{\rm vir}$, the net contribution in each shell is only 12--13 per cent of the sum of the absolute contributions. Moreover,
$J_{\parallel,{\rm shell}}/(M_{\rm vir}R_{\rm vir}V_{\rm vir})=(M_{\rm shell}/M_{\rm vir})\langle\widetilde{j}_\parallel\rangle_{\rm shell}$, so the particles beyond $R_{\rm vir}$ can have large specific angular momenta while contributing little to the total because those shells contain little mass.

\bibliographystyle{mn2e}
\bibliography{oja_template}

\end{document}